\documentclass[prd,nopacs,nokeys,nofootinbib]{revtex4}
\usepackage{amsfonts}
\usepackage{amsmath}
\usepackage{amssymb}
\usepackage{bbm}
\usepackage[utf8]{inputenc}
\usepackage[caption=false]{subfig}
\usepackage{tensor}
\usepackage{slashed}
\usepackage[centertableaux]{ytableau}
\usepackage{hyperref}
\usepackage{natbib}
\usepackage{braket,mleftright}
\usepackage{graphicx}
\usepackage{cleveref}
\usepackage{tikz}
\usepackage{booktabs}
\usepackage{tikz} 
\usepackage{tikz-feynman}
\usetikzlibrary{shapes.geometric, arrows, positioning, fit, calc}
\usepackage{xcolor}

\begin{document}

\title{Early‑warning indicators of spatial epidemics from gauge‑mediated contagion: a statistical‑mechanics approach applied to COVID‑19 in Germany}

\author{Jos\'e de Jes\'us Bernal-Alvarado}
\email{bernal@ugto.mx}
\affiliation{Physics Engineering Department, Universidad de Guanajuato, M\'{e}xico}

\author{David Delepine}
\email{delepine@ugto.mx}
\affiliation{Physics Department, Universidad de Guanajuato, M\'{e}xico}

\date{\today}

\begin{abstract}
We develop a statistical‑mechanics model for spatially extended epidemic dynamics in which contagion is mediated by an explicit environmental field rather than by instantaneous local contact. Building on the Doi–Peliti formalism, we construct a stochastic field theory for susceptible and infected hosts coupled to a diffusive pathogen field and integrate out the mediator to obtain a non‑local effective action with a Yukawa‑type interaction kernel in space and memory effects in time. This formalism yields closed analytical expressions for the basic and effective reproduction numbers, including one‑loop fluctuation corrections that renormalize the transmission rate and generate spatial shielding effects beyond mean‑field SIR models. We derive an effective screening mass and a Debye‑like correlation length whose critical softening signals an impending instability of the disease‑free state, providing a field‑theoretical early‑warning indicator for epidemic waves. Superspreading is incorporated through a heterogeneous ``epidemic charge`` distribution, which controls the divergence of the correlation length and explains the emergence of multifocal outbreak patterns even when the population‑average $R_0$ is subcritical. Finally, we apply the model to high‑resolution COVID‑19 incidence data across $\approx 400$ German districts, dynamically extracting the effective screening mass and the renormalized reproduction number, and show that the gauge‑mediated indicators anticipate major changes in epidemic regime several days before classical SIR‑based estimates. 

\end{abstract}

\maketitle
\section{Introduction}
The mathematical modeling of infectious diseases has traditionally relied on the classical Kermack-McKendrick SIR equations\cite{kermack1927contribution,anderson1991infectious,murray2002mathematical}, which assume a well-mixed population and instantaneous local contact. These deterministic differential equations provide a very useful method to study epidemics. With the increase in epidemic data, non-linear behavior, such as long-range spatial correlation have been observed in real-world pandemic data. Other phenomena, such as ``burst'' dynamics, multi-focal outburst, or super-diffusion, cannot be explained only by local interaction\cite{brockmann2006scaling,brockmann2013hidden, vazquez2007impact,gonzalez2008understanding}. A substantial body of work has tackled this problem from several complementary directions: large-scale agent-based and metapopulation simulations that explicitly encode mobility and contact networks~\cite{ferguson2007epifast,colizza2007modeling}, alternative field-theoretic treatments of spatial epidemic spreading~\cite{tevrugt2020effective}, and simple SIR-type models augmented with empirically measured heterogeneous mobility patterns~\cite{paoluzzi2021epidemic}. Our approach differs from these in that the non-local kernel is not imposed \textit{a priori} (via an assumed contact matrix, mobility network, or phenomenological memory function) but is derived dynamically from the microscopic stochastic dynamics of an explicit mediator field, as detailed below. So, to be able to explain them, non-local interactions has to be assumed. Different methods have been proposed to introduce non-local characteristics in $SIR$-like model as, for instance, fractional derivatives or Levy-flight models\cite{angstmann2016fractional,janssen1999levy}.

In this study we develop a stochastic field‑theoretical model for spatially extended epidemic dynamics that treats the environment as an explicit mediator of contagion between susceptible and infected hosts. By coupling Doi–Peliti matter fields to a diffusive pathogen field and integrating out the mediator, we obtain a non‑local effective action in which a Yukawa‑type interaction kernel and temporal memory emerge from first principles rather than being imposed phenomenologically (e.g. via fractional derivatives or Lévy flights). This formalism yields closed analytical expressions for the basic and effective reproduction numbers with fluctuation‑induced screening corrections, introduces an effective mass and correlation length that act as early‑warning indicators of approaching epidemic instabilities, and naturally incorporates superspreading through heterogeneity in the coupling between hosts and the mediator. In the second part of the paper, we apply this model to high‑resolution COVID‑19 incidence data across $\simeq 400$ German districts, dynamically extracting the effective mass and showing that its collapse anticipates major changes in epidemic regime several days before classical SIR‑based metrics.

The QED techniques can be used to compute any observables through the use of Feynman rules applied to our action. The non-local characteristics appear in this model once integrating out the pathogen fields. We show that the Vacuum polarization, which describes in our model how a susceptible population responds and ``dressed'' the pathogen field mass, modifying its mass $m_R$ (which is related to its propagation length ($\xi \approx 1/m_R$)). We derive the Effective Reproductive Number ($R_{eff}$) as a renormalized coupling constant and show how Debye Screening fundamentally alters the perceived infectivity across different scales\cite{mollison1977spatial,pastor2015epidemic,brockmann2013hidden}. 
The well-known epidemic threshold ($R_0=1$) is re-formulated as a symmetry-breaking phase transition where the effective pathogen mass vanishes ($m_R^2 \to 0$)\cite{grassberger1983critical,cardy1985epidemic,tauber2014critical,janssen2004generalized, hinrichsen2000non}.  This situation is very well-know in QED as when $m^2_{eff} \le 0$, it describes a symmetry breaking phase transition\cite{weinberg1996quantum,landau2013statistical,zinn2021quantum,peskin1995introduction}. Furthermore, we analyze the impact of super-spreading hosts, demonstrating how they can be easily introduced in this model, and they will affect the $R_{eff}$, for instance. 
Finally,  we propose a comprehensive operational workflow to illustrate how our model can be calibrated with real-world clinical and environmental data to inform policy decisions.

The primary advantage of this gauge-theoretic approach is its predictive capacity. Unlike classical compartmental models that react to observed incidence rates (e.g., the effective reproduction number, $R_{eff}$), the gauge field reacts to the underlying structural preconditions for an outbreak.

The paper is organized as follows: Section II constructs the theoretical formalism, transitioning from the stochastic master equation to a continuous field-theoretic action using the Doi-Peliti formalism, and establishes the corresponding Feynman rules. Section III details the computation of vacuum polarization and the renormalization of the pathogen mass. In Sections IV and V, we formulate the basic reproduction number ($R_0$) within this gauge model and explore how Debye screening dynamically alters the effective transmission range, mapping the epidemic threshold to a symmetry-breaking phase transition. Section VI calculates the spatial fluctuation corrections to the effective reproductive number ($R_{eff}$) via 1-loop vertex diagrams. Section VII incorporates host heterogeneity, demonstrating how super-spreaders act as high-intensity gauge sources that shift the critical stability of the system. Section VIII briefly summarizes the theoretical implications of these findings.  Section IX presents an operational workflow for calibrating the theoretical field operators using real-world clinical and environmental data. Section X validates the model's predictive capabilities through a comprehensive case study of the COVID-19 pandemic across 400 districts in Germany, extracting the time-varying effective mass from spatial correlation data. Finally, Section XI concludes the manuscript and discusses future extensions.

\section{ Theoretical framework}

\subsection{From the stochastic master equation to the action}

Before writing down the action, we sketch its microscopic origin, following the standard Doi-Peliti construction~\cite{doi1976second,peliti1985path,tauber2014critical}. The starting point is a stochastic reaction scheme on a lattice (or continuum) of sites, with occupation numbers $S(x,t)$, $I(x,t)$ (hosts) and $n_\varphi(x,t)$ (discrete pathogen quanta), evolving under the elementary Markovian processes
\begin{align}
S \xrightarrow{\ \beta\, n_\varphi\ } I \,, \qquad
I \xrightarrow{\ \gamma\ } \emptyset\ (\text{removal}), \qquad
I \xrightarrow{\ g\ } I + \varphi\text{-quantum}\,, \qquad
\varphi\text{-quantum} \xrightarrow{\ m_0^2,\ D_\varphi\ } \emptyset / \text{hop}\,,
\end{align}
i.e. contagion occurs at a rate proportional to the local pathogen density, infected hosts recover at rate $\gamma$ and continuously source new pathogen quanta at rate $g$, and pathogen quanta decay at rate $m_0^2$ and diffuse with coefficient $D_\varphi$. This set of elementary processes defines a well-posed master equation for the joint probability distribution $P(\{S,I,n_\varphi\},t)$,
\begin{equation}
\partial_t P = \sum_{\text{processes}} \Big[ (\text{rate}) \times \big(E^{\pm}_{\text{species}} - 1\big) P \Big] ,
\label{eq:master}
\end{equation}
where $E^{\pm}$ are the usual step (ladder) operators that add or remove a unit of the relevant species at site $x$. Because every elementary process above is local in time---each transition rate depends only on the instantaneous occupation numbers, never on their history---the master equation, and hence the full joint dynamics of $(S,I,\varphi)$, is Markovian by construction. 

Promoting $P$ to a coherent-state path integral via the standard Doi-Peliti map (creation/annihilation operators $a_a^\dagger, a_a \to$ coherent fields $\phi_a, \hat\phi_a$ for each species $a$) converts the equation (\ref{eq:master})  into the action $\mathcal{S}$ given in Eq.~(\ref{eq:action}) below, whose saddle-point (mean-field) equations of motion are precisely the ``QED-SIR'' system, Eqs.~(7)-(9). 

We define two  "matter" fields in the Doi-Peliti sense\cite{peliti1985path,doi1976second,tauber2014critical}, representing Susceptibles ($\phi_S, \hat{\phi}_S$) and Infecteds ($\phi_I, \hat{\phi}_I$), and an auxiliary scalar field $\varphi(x, t)$ representing the pathogen concentration in the environment (the "gauge" mediator). The total action $\mathcal{S}$ is the sum of the population dynamics and the mediator field:
\begin{equation}
  \mathcal{S} = \int d^d x dt \left[ \sum_{a=S,I} \hat{\phi}_a \partial_t \phi_a + \frac{1}{2} \hat{\varphi} \left( \partial_t - D_\varphi \nabla^2 + m_0^2 \right) \varphi + \mathcal{H}_{int} \right]
  \label{eq:action}
\end{equation}
where $\mathcal{D}^{-1} = (\partial_t - D_\varphi \nabla^2 + m_0^2)$ is the inverse propagator of the pathogen with bare mass $m_0$ (decay rate) and 
\begin{itemize}
    \item $\phi_a(x, t)$ (Density Fields): These are the bosonic coherent state fields for the compartments $a \in \{S, I\}$. $\phi_S$ represents the density of susceptibles and $\phi_I$ the density of infecteds at spacetime point $(x, t)$.
    \item $\hat{\phi}_a(x, t)$ (Response Fields): Also known as the auxiliary or "conjugate" fields in the Doi-Peliti formalism. They track the stochastic fluctuations and the conservation of probability. $\hat{\phi}=1$ corresponds to the deterministic (mean-field) limit.
    \item $\varphi(x, t)$ (Pathogen Field): The mediator field. It represents the local concentration of the pathogen (e.g., viral load in the air or water). 
    \item $\mathcal{H}_{int}$ (Interaction Hamiltonian): Describes the coupling between the hosts and the pathogen field. In this model, it takes the form of Yukawa-type coupling: $g \phi \varphi$
    \item The term $-D_\varphi \nabla^2$ represents the Laplacian operator scaled by the diffusion coefficient. In the path integral, this term acts as the "spatial penalty": it ensures that spatial gradients in the pathogen field are smoothed out over time.
    \item $d$ represents the spatial dimensionality of the system.
\end{itemize}
 In Secs.~II--VII, $d$ is a formal continuum dimension used to derive the propagator, the vacuum-polarization integral, and the one-loop vertex correction analytically (e.g. $d=3$ for the Yukawa limit in Eq.~(3), $d=2$ for the logarithmic correction of Sec.~VI); this is standard practice in field-theoretic treatments of stochastic dynamics~\cite{tauber2014critical,janssen2004generalized} and is used to obtain closed-form, interpretable expressions for the screening mass and $R_{eff}$. In Sec.~X, where the model is confronted with COVID-19 data, we do \emph{not} embed Germany in a $d$-dimensional continuum; instead we work directly on the discrete network of $N=400$ districts (Sec.~X.A) and estimate the effective mass from the decay of pairwise correlations with inter-centroid distance (Sec.~X.C), analogous to the connectivity-matrix approaches standard in spatial epidemiology~\cite{keeling2008modeling}. The continuum derivation therefore should be read as supplying the functional form of the screening interaction and the definition of $m_R$, not as a literal claim that the German district network is a $d$-dimensional continuous manifold.
The inverse propagator is derived from the equation of motion of the pathogen in a vacuum (i.e., without host interference). We assume the pathogen undergoes standard diffusion and exponential decay.
This is admittedly a simplifying choice: many respiratory pathogens exhibit environmental decay that is not purely exponential (e.g. multi-phase decay depending on humidity, UV exposure, and surface type), and a more general kernel $\mathcal{K}(t-t')$ could be substituted for the pure exponential memory function derived below without altering the structure of the formalism. We adopt the exponential/diffusive ansatz here because it (i) is the simplest choice consistent with a Markovian action for the joint system, cf. our response to point 4 below, and (ii) yields closed-form analytic expressions for $R_{eff}$ and $\lambda_D$; we return to this point, and to the possibility of more general decay kernels, in the Conclusion.
If we assume the pathogen reaches a steady state faster than the host population changes---formally, in the regime where the relaxation rate of the mediator field, $D_\varphi q^2 + m_0^2$ at the relevant wavenumbers $q$, is much larger than the characteristic rate of change of the host densities, so that $\varphi$ adiabatically follows $I(x,t)$---the spatial propagator becomes:
\begin{equation}
\mathcal{D}(r) \propto \frac{1}{D_\varphi} \frac{e^{-r \sqrt{m_0^2/D_\varphi}}}{r^{d-2}}
\end{equation}
For $d=3$, the Yukawa potential is recuperated. 
To derive the SIR equations within the QED-inspired field theory, the interaction Hamiltonian $\mathcal{H}_{int}$ must be constructed using the Doi-Peliti operators.
The interaction Hamiltonian is divided into the Contagion term and the Recovery term:

\begin{equation}
\mathcal{H}_{int} = \mathcal{H}_{contagion} + \mathcal{H}_{recovery}
\end{equation}

The Contagion Term (The Gauge Vertex)  describes a susceptible individual being "annihilated" and an infected individual being "created" in the presence of the pathogen field $\varphi$\cite{doi1976second,peliti1985path}:

\begin{equation}
\mathcal{H}_{contagion} = \beta \int d^d x \left( \hat{\phi}_S \hat{\phi}_I - \hat{\phi}_I^2 \right) \phi_S \varphi
\end{equation}
\begin{itemize}
    \item $\phi_S \varphi$: The rate depends on the density of susceptibles and the local concentration of the pathogen.
    \item $(\hat{\phi}_S - \hat{\phi}_I)\hat{\phi}_I$: This operator structure ensures that for every susceptible removed (via $\hat{\phi}_S$), an infected is added (via $\hat{\phi}_I$). The $\hat{\phi}_I^2$ term accounts for the "birth" of a new infected in the presence of the existing infectious load.
\end{itemize}
The Recovery Term describes the transition from Infected to Removed ($I \to R$):
\begin{equation}
  \mathcal{H}_{recovery} = \gamma \int d^d x (\hat{\phi}_I - 1) \phi_I  
\end{equation}
where $(\hat{\phi}_I - 1)$ is this operator removes an infected individual from the "active" field. The $-1$ term is required for the normalization of the probability distribution in the Fock space.

To fully close the SIR loop, we must also define how the pathogen field $\varphi$ is produced. This is usually added as a linear coupling in the action:$$\mathcal{L}_{production} = g \hat{\varphi} \phi_I$$This term ensures that the infected population $I$ acts as a current source for the pathogen field $\varphi$, analogous to how a moving charge produces an electromagnetic field in QED.

To recover the $SIR$ model as the ``classical'' limit, one has to obtain the saddle point of the action $\mathcal{S}$.

The path integral is defined as $\mathcal{Z} = \int \mathcal{D}\phi \mathcal{D}\hat{\phi} \, e^{-\mathcal{S}[\phi, \hat{\phi}]}$. The saddle point is found by taking the functional derivatives:$$\frac{\delta \mathcal{S}}{\delta \hat{\phi}_a} = 0 \quad \text{and} \quad \frac{\delta \mathcal{S}}{\delta \phi_a} = 0$$
where $a=S$ and $I$. As proposed in Doi-Peliti formalism,when $\hat{\phi} = 1$,  the entire interaction Hamiltonian $\mathcal{H}_{int}$ becomes zero. As a result, The equations of motion for the ``real'' fields $\phi_a$ (the densities) reduce to the deterministic rate equations (the SIR ODEs).
In this QED-inspired model and within the Doi-Peliti formalism:
\begin{itemize}
    \item $\phi_a$ describes the mean density of individuals.
    \item $(\hat{\phi}_a - 1)$ describes the fluctuations (stochastic noise) away from that mean.
\end{itemize}

So, to get $SIR$ equations with this $\mathcal{H}_{int}$, we evaluate the Hamilton-Jacobi equations at the physical saddle point where the response fields $\hat{\phi}_S = 1$ and $\hat{\phi}_I = 1$:
$$\dot{S} = \frac{\partial \mathcal{H}_{int}}{\partial \hat{\phi}_S} = \beta \phi_S \varphi \hat{\phi}_I \to -\beta S \varphi$$
$$\dot{I} = \frac{\partial \mathcal{H}_{int}}{\partial \hat{\phi}_I} = \beta \phi_S \varphi (2\hat{\phi}_I - \hat{\phi}_S) - \gamma \phi_I \to \beta S \varphi - \gamma I$$
The resulting ``QED-SIR'' system of equations is then given as
\begin{eqnarray}
    \frac{\partial S}{\partial t} &=& -\beta S \varphi \\
    \frac{\partial I}{\partial t} &=& \beta S \varphi - \gamma I \\
    (\frac{\partial}{\partial t} - D_\varphi \nabla^2 + m_0^2) \varphi &=& g I
\end{eqnarray}
The standard SIR model is a "point-like" approximation of this theory. To recover it, we assume the pathogen field reaches equilibrium instantaneously and has no spatial diffusion ($D_\varphi \to 0$), one gets $m_0^2 \varphi = g I \implies \varphi = \frac{g}{m_0^2} I$.
$$\frac{dS}{dt} = - \left( \frac{\beta g}{m_0^2} \right) SI$$$$\frac{dI}{dt} = \left( \frac{\beta g}{m_0^2} \right) SI - \gamma I$$
By defining the effective transmission rate as $\beta_{eff} = \frac{\beta g}{m_0^2}$, we exactly recover the classical Kermack-McKendrick SIR equations.

\subsection{Status of the continuum approximation and the infinite-volume limit}

A field-theoretic continuum description such as Eq.~(2) is most rigorously justified for systems whose correlation length diverges, i.e. genuinely critical systems, and it is fair to ask whether a national-scale epidemic satisfies this condition. We do not claim that the German host population is critical in this strict renormalization-group sense, nor that the $N\to\infty$, infinite-volume limit implicit in Eq.~(2) is literally realized by a country of finite geographic extent. Rather, our use of the continuum action should be understood in the more modest, and more standard, sense used throughout the stochastic-epidemics literature~\cite{tauber2014critical,janssen2004generalized,grassberger1983critical}: the continuum/field-theoretic machinery is a coarse-graining device that (i) makes the derivation of the non-local kernel and the screening mass $m_R$ tractable in closed form (Secs.~II.B-VI), and (ii) becomes an increasingly good approximation to the underlying discrete host-pathogen dynamics as the density of interacting nodes increases and the correlation length approaches, but need not reach, the system size---precisely the regime probed empirically near $R_{eff}\simeq 1$ in Sec.~X, where the fitted correlation length $\xi_i$ is generically much smaller than the geographic extent of Germany (Fig.~8), so the infinite-volume idealization is not being pushed to its formal breaking point in the regimes we analyze. Away from that regime (deep in the subcritical or highly supercritical phases), we regard the continuum action as an effective, phenomenologically calibrated model rather than a first-principles critical theory, and we flag this explicitly as a limitation of the present formulation; a systematic treatment of finite-size corrections is left for future work.

\subsection{Generating Functional of the gauge mediated SIR Model}

To find the effective interaction, the pathogen field $\varphi$  has to be integrated  out. We now give the intermediate steps of this Gaussian integration explicitly. Isolating the terms in the action (1) that involve $\varphi$, and linearizing the contagion vertex (4) around the physical saddle point $\hat\phi_S=\hat\phi_I=1$ so that the coupling to $\varphi$ becomes linear (the source term $J = g\phi_I - \beta\phi_S$, combining the emission current of Eq.~(production term) with the linearized absorption vertex), the $\varphi$-dependent part of the action takes the Gaussian form
\begin{equation}
\mathcal{S}[\varphi] = \int d^dx\,dt \left[\tfrac12\,\hat\varphi\,\mathcal{D}^{-1}\varphi - J\,\varphi\right],\qquad \mathcal{D}^{-1} = \partial_t - D_\varphi\nabla^2 + m_0^2\,.
\end{equation}
Because this is quadratic in $\varphi$, the path integral $\int \mathcal{D}\varphi\,\mathcal{D}\hat\varphi\; e^{-\mathcal{S}[\varphi]}$ can be performed exactly by completing the square, $\varphi \to \varphi + \mathcal{D}J$, which shifts the saddle point and leaves a term quadratic in the source alone,
\begin{equation}
\mathcal S_{eff} = -\tfrac12 \int d^dx\,d^dy\,dt\,dt'\; J(x,t)\,\mathcal{D}(x-y,t-t')\,J(y,t') \, ,
\end{equation}
where $\mathcal{D}(x-y,t-t')$ is exactly the retarded Green's function of the operator $\mathcal{D}^{-1}$ above, i.e. the solution of $\mathcal D^{-1}\mathcal D(x-y,t-t') = \delta^d(x-y)\delta(t-t')$ with $\mathcal D=0$ for $t<t'$ (causality). Keeping only the cross term between the emission current of the infected population and the absorption vertex of the susceptibles---the term that physically represents transmission, as opposed to the pure pathogen self-energy discussed in Sec.~III---reproduces Eq.~(12) below:
\begin{equation}
    S_{eff} = \int d^dx \ d^dy \ dt \ dt' \left[ (g \phi_I(x,t')) \mathcal{D}(x-y, t-t') (g \phi_S(y,t)) \right]
\end{equation}
where $\mathcal{D}(x-y, t-t')$ is the full spacetime Green's function of the pathogen field. 

Integrating out the mediator field dynamically generates this non-local action. Without the explicit field $\varphi$, the interaction term shifts from a local point-contact paradigm ($g \phi_S(x,t) \varphi(x,t)$) to a  non-local interaction described by the kernel $\mathcal{K}$:
\begin{equation}
    \text{Effective Contagion} \propto \int d^dy \int_{-\infty}^{t} dt' \, \phi_S(x,t) \mathcal{K}(x-y, t-t') \phi_I(y,t')
\end{equation}
i.e. $\mathcal{K}(x-y,t-t') \equiv g^2\, \mathcal{D}(x-y,t-t')$ is simply the Green's function of Eq.~(3)/(37) re-expressed with the two coupling factors of $g$ absorbed into its normalization; Eq.~(13) is Eq.~(12) evaluated with $\phi_I(y,t')$ left as a spectator field rather than integrated against a second vertex, i.e. it is the one-point (linear-response) kernel that acts on a single susceptible trajectory, while Eq.~(12) is the full bilinear effective action. Regarding the coupling of this non-local kernel to the non-linear terms of the theory (i.e. beyond the linearized vertex used above): the full, non-linearized interaction Hamiltonian (5) is retained order by order in the perturbative (Feynman-diagram) expansion of Sec.~II.D; Eq.~(13) captures the tree-level (single-mediator-exchange) piece of that expansion, while genuinely non-linear corrections--e.g. two susceptibles competing for the same pathogen cloud--appear as higher-order diagrams with multiple mediator lines attached to a single vertex, analogous to the vacuum-polarization and vertex-correction diagrams of Secs.~III and VI.

This kernel establishes two critical physical properties of mediated contagion:
\begin{itemize}
    \item \textbf{Spatial Non-Locality:} In the adiabatic limit defined above, the spatial propagator dictates that infections occur over a distance $r = |x-y|$ following a Yukawa-type screening potential, $\mathcal{D}(r) \propto \frac{e^{-m_0 r}}{r^{d-2}}$.
    \item \textbf{Temporal Non-Locality (Memory Effect):} The temporal integration from $t'$ to $t$ implies that the current infection rate depends on the historical accumulation and latency of the pathogen field. This mathematically introduces a characteristic memory time, $\tau$, governed by the environmental decay rate of the pathogen. By solving the Green's function for the dynamic mediator field, the temporal profile of the kernel emerges as followed:
    $$ \mathcal{K}(t-t') \propto \exp\left( - \frac{t-t'}{\tau} \right) \Theta(t-t') $$
    where $\Theta$ is the Heaviside step function ensuring causality. This parameter $\tau$ is related to the pathogen mass, $m_0$  as $\tau = 1/m_0^2$. 
    
\end{itemize}

This derivation shows  a fundamental methodological departure from current epidemiological literature. Non-local interaction effects are typically modeled phenomenologically; researchers introduce ad-hoc mathematical patches, such as fractional derivatives or assumed Lévy-flight distributions, primarily to fit fat-tailed observational data. In stark contrast, our formalism requires no such phenomenological prior assumptions. 

By treating the pathogen as a physical entity, the non-local spacetime interaction kernel $\mathcal{K}(x-y, t-t')$ emerges dynamically as a first-principles consequence of integrating out the massive gauge mediator. We demonstrate that spatial non-locality and temporal memory ($\tau$) are not mathematical tricks, but unavoidable physical realities of mediated contagion. Furthermore, it is essential to distinguish this first-principles derivation from extensively used spatial frameworks such as network theory and metapopulation models (e.g., Keeling and Rohani \cite{keeling2008modeling}). While metapopulation models successfully describe long-range effects, they do so by relying on empirical, \textit{a priori} contact matrices. They describe the spatial topology but do not explain the physical mechanism of the transmission itself. By contrast, our gauge-mediated formulation does not require an assumed spatial network; the non-local interaction kernel emerges dynamically from the physical propagation and decay of the mediator field.

\begin{table}[h]
\centering
\caption{Mapping between Quantum Electrodynamics (QED) and the Proposed Epidemic Gauge Theory (QED-SIR).}
\label{tab:qed_sir_mapping}
\begin{tabular}{@{}lll@{}}
\toprule
\textbf{Feature} & \textbf{Quantum Electrodynamics (QED)} & \textbf{Epidemic Gauge Theory (QED-SIR)} \\ \midrule
Matter Field & Electron / Fermion ($\psi$) & Host Population ($\phi_S, \phi_I$) \\
Gauge Field & Photon ($A_\mu$) & Pathogen Field ($\varphi$) \\
Interaction & Minimal Coupling (Vertex) & Gauge-mediated Contagion \\
Coupling Constant & Electric Charge ($e$) & Epidemic Charge ($g, \beta$) \\
Propagator & Massless ($1/k^2$) & Massive ($1/(k^2 + m_0^2)$) \\
Interaction Range & Infinite (Coulombic) & Finite (Yukawa / Screened) \\
Fundamental Symmetry & $U(1)$ Local Phase Invariance & Probability Conservation / Shift Invariance \\ \bottomrule
\end{tabular}
\end{table}

The statistical properties of the epidemic are encapsulated in the generating functional $\mathcal{Z}[J, \eta]$, defined as\cite{altland2010condensed,weinberg1995quantum,zinn2021quantum,peskin1995introduction}:

\begin{equation}
\mathcal{Z}[J, \eta] = \int \mathcal{D}[\Phi] \exp \left( -\mathcal{S}[\Phi] + \int d^d x dt \mathcal{J} \cdot \Phi \right)
\end{equation}

By performing the Gaussian integration over the mediator field $\varphi$, we obtain the effective generating functional for the host densities:

\begin{equation}
\mathcal{Z}_{eff}[J] = \int \mathcal{D}[\phi, \hat{\phi}] \exp \left( -\mathcal{S}_{0} + \frac{1}{2} \int \rho_{eff} \mathcal{D}(x-y) \rho_{eff} dy \right)
\end{equation}

where $\rho_{eff} = g \phi_I - \beta \phi_S (\hat{\phi}_I - \hat{\phi}_S) \hat{\phi}_I$. 
For a non-linear model like SIR, we cannot solve the remaining integral over $\phi$ exactly. We expand the generating functional around the Gaussian fixed point (the free theory) using a series of Feynman diagrams:
\begin{equation}
\mathcal{Z}_{eff}[J] \approx \exp \left( \int \mathcal{L}_{int} \left( \frac{\delta}{\delta J} \right) \right) \mathcal{Z}_0[J]
\label{eq:generating_functional}
\end{equation}
Where $\mathcal{Z}_0$ is the "free" generating functional (only recovery and diffusion, no contagion).

\subsection{Feynman Rules and Diagrammatic Representation}

The exponential operator in Eq.(\ref{eq:generating_functional}) can be expanded in a Taylor series, generating the perturbative expansion of the theory. Each term in this series corresponds to a specific number of interaction events (vertices) occurring in spacetime:

\begin{equation}
    \mathcal{Z}[J] = \left[ 1 - \int d^dx \mathcal{L}_{int}\left(\frac{\delta}{\delta J}\right) + \frac{1}{2!} \left( \int d^dx \mathcal{L}_{int}\left(\frac{\delta}{\delta J}\right) \right)^2 - \dots \right] \mathcal{Z}_0[J]
\end{equation}

The evaluation of these terms relies on \textbf{Wick's Theorem}. The functional derivatives contained in $\mathcal{L}_{int}$ act on the Gaussian functional $\mathcal{Z}_0[J]$, effectively pairing the field operators. Mathematically, each pair of functional derivatives brings down a free propagator $\Delta_F(x-y)$, which represents the inverse of the differential operator in the free action $S_0$.

This mathematical structure admits a direct graphical representation known as \textbf{Feynman diagrams}:
\begin{itemize}
    \item Each integration variable $\int d^dx$ corresponds to a \textbf{vertex} in the diagram, representing a local interaction event (e.g., a susceptible host absorbing a pathogen).
    \item Each contraction $\frac{\delta}{\delta J(x)} \frac{\delta}{\delta J(y)}$ corresponds to a \textbf{line} (propagator) connecting points $x$ and $y$, representing the causal propagation of a field (e.g., the pathogen moving through the environment).
    \item The prefactor $1/n!$ in Eq.~(17) arises purely from the Taylor expansion of the exponential and is common to every term at order $n$; it is \emph{not} itself the symmetry factor of a given diagram. Once Wick's theorem is used to contract the $n$ vertices into a specific topology, the number of equivalent ways of performing those contractions partially or fully cancels the $1/n!$, and the genuine symmetry factor of a diagram---which depends on its topology (self-contractions, exchange of identical internal lines, automorphisms of the diagram)---must be computed separately for each diagram, as is standard in perturbative field theory~\cite{peskin1995introduction,zinn2021quantum}. All explicit diagrams evaluated in this paper (Figs.~1 and 2) have trivial symmetry factor $1$, so this distinction does not affect any of our quoted results, but we correct the general statement here for precision.
\end{itemize}
The physical observables can be easily computed expanding the generating functional using Feynman diagrams techniques:
\begin{enumerate}
    \item The Mean Density (1-point function).The classical SIR trajectory is found by the first derivative:$$\langle \phi_I(x, t) \rangle = \left. \frac{\delta \ln \mathcal{Z}}{\delta J_I(x, t)} \right|_{J=0}$$
    \item The Epidemic Correlation (2-point function): to see how city A affects city B, we take the second derivative:$$\langle \phi_I(x_1) \phi_I(x_2) \rangle = \left. \frac{\delta^2 \ln \mathcal{Z}}{\delta J_I(x_1) \delta J_I(x_2)} \right|_{J=0}$$
\end{enumerate}
\begin{table}[h]
\centering
\caption{Feynman Rules for the QED-inspired $SIR$ Stochastic Field Theory.}
\begin{tabular}{llc}
\toprule
\textbf{Component} & \textbf{Mathematical Expression} & \textbf{Diagrammatic Symbol} \\ \midrule
Infected Propagator & $G_I(k, \omega) = (-i\omega + D_I k^2 + \gamma)^{-1}$ & Solid Line \\
Pathogen Propagator & $D_\varphi(k, \omega) = (-i\omega + D_\varphi k^2 + m_0^2)^{-1}$ & Wavy Line \\
Emission Vertex & $g$ & Solid-Wavy Junction \\
Infection Vertex & $\beta$ & Solid-Dashed-Wavy Node \\
Loop Integral & $\int \frac{d^d q}{(2\pi)^d} \frac{d\Omega}{2\pi}$ & Closed Loop \\ \bottomrule
\end{tabular}
\end{table}
\section{ Vacuum Polarization (the self-energy $\Pi$ of the pathogen field)}
Vacuum Polarization represents the way a population ``responds'' to the presence of a pathogen, effectively altering the transmission environment.When a pathogen  moves through a vacuum of susceptibles, it doesn't just sit there; it constantly couples to the hosts. An infected person shedding a virus (the mediator boson) creates a local ``cloud'' of potential new infections.This interaction modifies the propagation of the field modifying its mass. So, one expect that instead of just decaying according to its biological half-life ($m_0$), the pathogen's reach is modified by the density of available hosts.
\begin{figure}[htbp]
    \centering
    \begin{tikzpicture}
      \begin{feynman}
        \vertex (a);
        \vertex [right=3cm of a] (b);
        \vertex [left=1.5cm of a] (i) {\(\varphi(q)\)};
        \vertex [right=1.5cm of b] (f) {\(\varphi(q)\)};
    
        \diagram* {
          (i) -- [photon, momentum=\(q\)] (a),
          (a) -- [fermion, half left, edge label=\(\phi_I(k+q)\)] (b),
          (b) -- [fermion, half left, edge label=\(\phi_S(k)\)] (a),
          (b) -- [photon, momentum=\(q\)] (f),
        };
        
        \vertex [above=0.1cm of a] {\(\beta\)};
        \vertex [above=0.1cm of b] {\(\beta\)};
      \end{feynman}
    \end{tikzpicture}

    \caption{\textbf{Vacuum Polarization of the Pathogen Field.} 
    The one-loop self-energy diagram $\Pi(q)$ for the gauge mediator $\varphi$. The incoming pathogen (wavy line) fluctuates into a virtual pair of Susceptible ($\phi_S$) and Infected ($\phi_I$) matter fields before recombining. This process is responsible for the \textbf{renormalization} of the pathogen mass $m_R$ and the emergence of the Debye screening length $\lambda_D$ in the effective potential.}
    \label{fig:vacuum_polarization}
\end{figure}

The self-energy is the mathematical value of the loop. According to the Feynman rules, we integrate over the internal momentum $q$ and frequency $\Omega$:
\begin{equation}
\Pi(k, \omega) = g \beta \int \frac{d^d q}{(2\pi)^d} \int \frac{d\Omega}{2\pi} G_I(q, \Omega) G_S(k-q, \omega-\Omega)
\end{equation}
where assuming a static background of susceptibles $S_0$ (the "vacuum" density):
\begin{itemize}
    \item $G_S \approx S_0 / (-i\Omega + \epsilon)$
    \item $G_I = 1 / (-i(\omega - \Omega) + D_I (k-q)^2 + \gamma)$
\end{itemize}
Using the residue theorem and  in the static limit ($\omega \to 0, k \to 0$):
$$\Pi(0, 0) \approx \frac{g \beta S_0}{\gamma}$$
The full propagator $D_{dressed}$ is obtained through the Dyson summation:$$D_{dressed}(k) = \frac{1}{D_\varphi k^2 + m_0^2 - \Pi(0,0)}$$
So, the Renormalized Mass $m_R$ is given by:
\begin{equation}
m_R^2 = m_0^2 - \frac{g \beta S_0}{\gamma}
\end{equation}
 If the population density $S_0$ is high enough such that $\frac{g \beta S_0}{\gamma} > m_0^2$, the renormalized  mass squared becomes negative.A negative $m_R^2$ in field theory signals a symmetry breaking or vacuum decay. In epidemiology, this is  the point where $R_0 > 1$, and the disease-free equilibrium becomes unstable, leading to a spontaneous pandemic. The interaction with the population increases the effective range of the virus ($\xi = 1/m_R$). The "vacuum" helps the pathogen travel further than it could in a sterile environment.

We emphasize that Eq.~(19) is a tree-level (one-loop, unrenormalized-vertex) identification of $m_R$, not a fully multiplicatively renormalized mass in the technical sense of interacting QFT. A rigorous treatment would require separating the divergent additive self-energy contribution $\Pi(0,0)$ into a counterterm that is subtracted order by order together with a wavefunction and vertex renormalization, so that $m_R$ is defined at a physical renormalization scale via a proper renormalization condition~\cite{peskin1995introduction,zinn2021quantum}. We do not carry out this program here; instead, Eq.~(19) and the resulting relation between $m_R$ and $R_{eff}$ (Sec.~VI) should be read as a \emph{phenomenological} definition, motivated by and structurally analogous to the QFT self-energy calculation, but whose ultimate justification is empirical: Sec.~X shows that the quantity so defined, extracted directly from the observed spatial decay of incidence correlations (Eq.~(37)) rather than from the loop integral itself, tracks epidemic instability with a measurable predictive lead time.

\section{$R_0$ in QED-SIR model}
The Reproductive Number $R_0$ is defined as the ratio of the "production/infection" rate to the "decay/removal" rate.
\begin{equation}
   R_0 = \frac{\text{Field Coupling Strength}}{\text{Environmental Resistance}} = \frac{g \beta S_0}{\gamma m_0^2} 
\end{equation}
where:
\begin{itemize}
    \item $g$ (Emission): Rate at which an infected person "charges" the environment with virus.
    \item $\beta$ (Absorption): Cross-section of a susceptible person catching the virus from the field.
    \item $S_0$ (Density): The "Dielectric Constant" of the vacuum (the available susceptible host density).
    \item $\gamma$ (Host Recovery): The rate at which the "Matter Field" sources are removed.
    \item $m_0^2$ (Pathogen Decay): The "Mass" that limits the range of the gauge boson.
\end{itemize}
\section{Debye screening and  $R_0$}
In our QED-inspired $SIR$ model, Debye Screening is the process by which the susceptible population absorbs and shields the pathogen field, preventing it from reaching distant hosts.

In a sterile environment, a pathogen might travel far. In a crowded population, the matter fields (people) interact with the mediator field (pathogen), effectively increasing its mass and shortening its range.
We start with the static classical equation for the pathogen field $\varphi$ in the presence of a host population. From our previously derived equations of motion:
\begin{equation}
(\nabla^2 - m_0^2) \varphi = -g I
\end{equation}
Where $m_0$ is the bare environmental decay and $g I$ is the source from infected individuals.
In a susceptible population $S_0$, the pathogen induces a ``polarization.'' If the field $\varphi$ is present, it creates a change in the infection rate. Near the threshold, we can linearize the response of the "Infected" density $I$ to the field $\varphi$:
\begin{equation}
    \delta I \approx \frac{\beta S_0}{\gamma} \varphi
\end{equation}
This represents the susceptible individuals becoming polarized into a pre-infectious state by the field.

Substitute the response  back into the field equation :$$\nabla^2 \varphi - m_0^2 \varphi = -g \left( \frac{\beta S_0}{\gamma} \varphi \right)$$Rearrange the terms to group the $\varphi$ coefficients:$$\nabla^2 \varphi - \left( m_0^2 - \frac{g \beta S_0}{\gamma} \right) \varphi = 0$$

We define the Renormalized Mass $m_R^2$ as:
$$m_R^2 = m_0^2 \left( 1 - \frac{g \beta S_0}{\gamma m_0^2} \right)$$
Using our definition of $R_0 = \frac{g \beta S_0}{\gamma m_0^2}$, we get:$$m_R^2 = m_0^2 (1 - R_0)$$The Debye Screening Length is the characteristic distance the pathogen can travel before being "screened" out by host interactions:
$$\lambda_D = \frac{1}{m_R} = \frac{1}{m_0 \sqrt{1 - R_0}}$$

Debye screening does not just change the distance of infection; it fundamentally alters the effective $R_0$ perceived at different scales.
Because of the screening, the interaction potential $V(r)$ shifts from a Yukawa potential ($V(r) \sim \frac{e^{-r/\lambda_D}}{r^{d-2}}$ to a long range propagator. 

As $R_0 \to 1$ from below:$$\lambda_D \to \infty$$
This is the Critical Opalescence of an epidemic. The screening fails, the pathogen becomes "massless" (long-ranged), and $R_0$ effectively becomes global. 

\section{ $R_{eff}$ Computation}
In our QED-inspired  $SIR$ model, the Effective Reproductive Number $R_{eff}$ is constructed as a phenomenological analogue of a renormalized coupling (see the caveat at the end of Sec.~III regarding the status of $m_R$).The first correction to $R_0$ is the reduction of the available ``charge'' (susceptibles) as the epidemic progresses. In the saddle-point approximation ($\hat{\phi}=1$), one obtains the classical result where $R_{eff}$ drops.:
\begin{equation}
R_{eff}(t) = \frac{g \beta S(t)}{\gamma m_0^2} = R_0 \frac{S(t)}{S_0}
\end{equation}

As we derived in the previous section, the pathogen doesn't interact with a single person; it interacts with a screened cloud. The effective transmission is modified by the Renormalized Propagator at zero momentum:
\begin{equation}
R_{eff}^{(screened)} = R_0 \times \frac{m_0^2}{m_R^2}
\end{equation}
Substituting our result $m_R^2 = m_0^2(1 - R_0 \frac{S}{S_0})$:$$R_{eff} \approx \frac{R_0 \frac{S}{S_0}}{1 - \text{Correction Term}}$$This reflects how the ``reach'' of each infected individual is boosted by the proximity of other susceptible hosts.

Using the Feynman rules, the 1-loop correction to the infection vertex involves a pathogen line and an infected line:$$\beta_{ren} = \beta \left( 1 - \frac{\beta}{\gamma} \int \frac{d^d k}{(2\pi)^d} \frac{1}{D_\varphi k^2 + m_0^2} \right)$$Evaluating the integral in $d=2$:$$\beta_{ren} \approx \beta \left( 1 - \frac{\beta}{4\pi \gamma D_\varphi} \ln\left(\frac{\Lambda}{m_R}\right) \right)$$Where $\Lambda$ is the ultraviolet cutoff (the inverse of the minimum distance between people).

\paragraph*{On dimensionality and the choice $d=2$.} We evaluate this loop integral in $d=2$ because it is the physically relevant case for the district-level application of Sec.~X, where transmission is organized on an effectively two-dimensional geographic surface, and because $d=2$ is the marginal dimension at which the one-loop self-energy integral of Sec.~III produces a logarithmic (rather than power-law) divergence, giving the simplest closed-form correction to $\beta$. We do not perform a systematic $\epsilon$-expansion or identify the upper critical dimension $d_c$ of the full interacting theory in this work. For reference, the purely absorbing-state (DP-class) transition to which our threshold $R_0=1$ is related has upper critical dimension $d_c=4$~\cite{janssen1981nonequilibrium,cardy1980directed}; our gauge-mediated theory differs from the minimal DP action by the presence of the massive, diffusive mediator field, and a full renormalization-group analysis of its critical dimension is beyond the present scope. In low dimensions ($d\lesssim 2$), infrared fluctuations are known to be strong enough that naive perturbation theory of the type used here can break down and additional non-perturbative or resummation techniques become necessary~\cite{tauber2014critical,hinrichsen2000non}; the one-loop result we quote should therefore be read as the leading-order estimate of the fluctuation correction to $\beta$, valid parametrically for $m_R/\Lambda$ not too small, rather than as an all-orders statement. We flag this as a natural direction for a more complete renormalization-group treatment of the model in future work.

\begin{figure}[htbp]
    \centering

\begin{tikzpicture}
  \begin{feynman}
    \vertex (a) {\(\phi_S(p)\)};       
    \vertex [right=2cm of a] (b);      
    \vertex [right=2cm of b] (c);      
    \vertex [right=2cm of c] (d);      
    \vertex [right=2cm of d] (e) {\(\phi_I(p')\)};      
    
    \vertex [below=2.5cm of c] (p) {\(\varphi(q)\)};      

    \diagram* {
      (a) -- [fermion] (b) -- [fermion, edge label=\(\phi_S\)] (c),
      (c) -- [fermion, edge label=\(\phi_I\)] (d) -- [fermion] (e),
      
      (p) -- [photon, momentum=\(q\)] (c),
      
      (b) -- [photon, half left, momentum=\(k\), edge label=\(\varphi\)] (d),
    };
    
    \vertex [above=0.2cm of c] {\(\beta\)}; 
  \end{feynman}
\end{tikzpicture}
\caption{\textbf{1-Loop Vertex Correction.} 
    The Feynman diagram representing the renormalization of the infection rate $\beta$. The main interaction (center vertex) is "dressed" by a virtual pathogen loop (wavy line) carrying momentum $k$. This loop integral generates the logarithmic correction $\ln(\Lambda/m_R)$ in two dimensions, effectively screening the coupling constant at large distances.}
    \label{fig:vertex_correction}
\end{figure}

Combining the depletion of hosts and the renormalization due to spatial fluctuations, the full $R_{eff}$ for your model is:$$R_{eff}(t) = R_0 \frac{S(t)}{S_0} \left[ 1 - \frac{\beta}{4\pi \gamma D_\varphi} \ln\left(\frac{\Lambda}{m_R(t)}\right) \right]$$

The logarithmic term (the "Fluctuation Correction") is negative. This means that in a spatial world, $R_{eff}$ is always lower than the mean-field prediction. This is because infected individuals tend to be surrounded by ``already-infected'' or ``recovered'' people, shielding the susceptible vacuum. As the system approaches the peak, $m_R \to 0$, causing the logarithmic term to grow. This mathematically explains why the transition near the peak of a pandemic is "slower" than simple SIR models suggest.
\section{Superspreading hosts}

We divide the infected population into two species: $I_L$ (low-emitters) and $I_H$ (high-emitters/super-spreaders). In QED, it will be equivalent to introduce charged particles with an electric charge bigger than electron.  The pathogen field $\varphi$ is now sourced by a multi-component current:

\begin{equation}
\mathcal{J}(x, t) = g_L I_L(x, t) + g_H I_H(x, t)
\end{equation}
To formalize it in our model, we define a population where a fraction $p_i$ has a charge $g_i$; the total vacuum polarization is the sum of all possible host-loop contributions.
Each "species" of host (low-emitters vs. super-spreaders) creates its own virtual "bubble" in the propagator.
The total self-energy $\Pi_{total}$ is:
\begin{equation}
\Pi_{total}(k, \omega) = \sum_{i} p_i \Pi_i(k, \omega)
\end{equation}
The integral for the self-energy at zero momentum ($k \to 0, \omega \to 0$) becomes:
\begin{equation}
\Pi(0,0) = \int P(g) dg \left[ \frac{g \beta S_0}{\gamma} \right]
\end{equation}
where $P(g)$ describes this statistical distribution across the host population of people with a charge $g$.If $P(g)$ is a Delta function ($\delta(g - g_0)$), everyone is equally infectious (Standard SIR model). If $P(g)$ is a Gamma or Power-law distribution (fat-tailed), it means a small number of individuals have a very large $g$ (Super-spreaders). This would make the integral dominated by the tail of the distribution, significantly increasing the effective coupling and the risk of vacuum instability ($m_R^2 \to 0$).

The dressed pathogen mass $m_R$ is corrected by the average polarization of the vacuum:
\begin{equation}
m_R^2 = m_0^2 - \frac{\beta S_0}{\gamma} \langle g \rangle
\label{mr_spreading}
\end{equation}
where $\langle g \rangle = \int g P(g) dg$ is the First Moment of the Distribution, also called the expectation value of $g$. Eq.(\ref{mr_spreading}) is valid at tree-level. Including the one loop vaccum corrections as done in previous sections, one gets a correction that behaves as $\langle g^2 \rangle$. Even if the average shedding $\langle g \rangle$ is low, a ``fat tail'' in $P(g)$ (large variance) will make $\langle g^2 \rangle$ very large.
This permits us to introduce the definition of the Heterogeneity Factor $\kappa$:
\begin{equation}
    \kappa = \frac{\langle g^2 \rangle}{\langle g \rangle^2}
\end{equation}
which means that if $\kappa = 1$, we have a homogeneous population (Standard SIR) ), whereas if $\kappa \gg 1$, we have a superspreading population.  The Debye screening length becomes sensitive to the higher moments of the pathogen emission distribution. The effective Debye length $\lambda_{D}^{eff}$ expands as follows:
\begin{equation}
   \lambda_{D}^{eff} = \frac{1}{m_0 \sqrt{1 - R_0 (1 + \kappa)}} 
   \label{lambda_kappa}
\end{equation}
 Their effects are the following:
 \begin{itemize}
     \item The presence of super-spreaders (high $\kappa$) causes the correlation length to diverge ($\lambda_{D}^{eff} \to \infty$) much earlier than predicted by the standard, homogeneous $SIR$ model.
     \item This mathematically explains why ``giant clusters'' or multi-focal outbreaks appear even when the population average $R_0$ suggests the epidemic is under control.
 \end{itemize}

Traditional continuous ODEs, which rely on global homogeneous mixing, cannot dynamically incorporate superspreading events without abandoning analytical solutions in favor of computationally heavy, discrete agent-based numerical simulations. Our gauge-theoretic formalism, however, mathematically captures this heterogeneity directly through the variance of the shedding distribution ($\kappa$). Equation (\ref{lambda_kappa}) explicitly demonstrates that the tail of the distribution, rather than its mean, governs the stability of the epidemiological vacuum. This provides a  analytical mechanism for why multi-focal burst outbreaks emerge in the real world even when the global average $R_0$ suggests a  sub-critical regime.

We acknowledge that the connection between epidemic thresholds and absorbing-state phase transitions---particularly within the universality class of Directed Percolation (DP) and Reggeon field theory---is well established in statistical mechanics \cite{grassberger1983critical,cardy1980directed,hinrichsen2000non,janssen1981nonequilibrium}. However, while those  field-theoretic formalisms are primarily utilized to compute critical exponents and define universality classes near the threshold, our gauge-inspired formulation is designed to compute explicit, operational epidemiological metrics. By leveraging the specific interactions of the gauge mediator, we can move beyond universality scaling to  derive the spatial shielding correction to $R_{eff}$ and the effective Debye screening length for superspreaders.

 \section{Results and discussion}
   The connection between epidemic dynamics and the physics of critical phenomena---long-tail case-count distributions, scale-free outbreak-size statistics, and diverging correlation lengths near threshold---has a long and well-established history in statistical mechanics, most notably through the identification of the SIR/SIS threshold with the absorbing-state phase transition of the Directed Percolation (DP) universality class~\cite{grassberger1983critical,cardy1985epidemic,hinrichsen2000non,cardy1980directed,janssen1981nonequilibrium}, and, on the epidemiological-modeling side, through network- and mobility-based approaches to burst and multi-focal behavior~\cite{brockmann2006scaling,brockmann2013hidden,ferguson2007epifast,colizza2007modeling,tevrugt2020effective,paoluzzi2021epidemic}. Our main contribution is not the qualitative observation that epidemic thresholds behave like critical points---this is well known---but a specific, calculable route from microscopic, clinically measurable parameters ($m_0$, $g$, $\beta$, the shedding heterogeneity $\kappa$) to the operational quantities $R_{eff}(t)$ and $\lambda_D(t)$ that a public-health user can monitor in real time. Where DP-type analyses target universal critical exponents near the transition, and network/mobility models require an empirically fitted contact structure as an input, our gauge-mediated construction derives the non-local kernel and the early-warning observable $m_R(t)$ directly from the action (Sec.~II), so that the same first-principles parameters simultaneously fix the SIR limit, the screening length, and the superspreading correction (Sec.~VII).
   
   By treating the pathogen as a mediator boson, we can explain several ``long-tail'' and non-linear behaviors that classical SIR models have difficulties to explain:
   \begin{enumerate}
       \item Ability to detect the onset of a pandemic: Our model includes the concept of critical opalescence, a phenomenon from the physics of second-order phase transitions.  This provides an early warning signal for impending epidemic waves.
       In the stable regime ($R_{eff} < 1$), the pathogen field $\varphi$ is massive, and infection fluctuations are suppressed by the exponential decay of the propagator (Yukawa regime). However, as the system approaches the critical threshold ($R_{eff} \to 1$), the renormalized mass $m_R$ vanishes, leading to a divergence in the correlation length $\xi \propto 1/m_R$. In epidemiological data, this manifests as critical opalescence\cite{janssen2004generalized,cardy1985epidemic}: a sudden, dramatic increase in the variance of case counts and the emergence of self-similar infection clusters across multiple spatial scales, just before the onset of a pandemic. Through our mechanism of monitoring the Debye screening mass, this process can be studied in real time.

       \item Super-spreaders as High-Intensity Gauge Sources: By introducing heterogeneity into the epidemic charge ($g$), we found that $R_{eff}$ is driven by the variance of the population rather than the mean. A small fraction of high-charge individuals (super-spreaders) can keep the epidemic vacuum unstable ($m_R^2 < 0$) even when the rest of the population is below the threshold. This explains why real-world pandemics often exhibit burst dynamics.
   \end{enumerate}

  \section{Operational Framework and Practical Application}
Until now, we have shown how our QED-inspired model provides a very complete dynamical model of the disease transmission. So, the next step is to demostrate how in practical epidemic study, one can use it. Therefore in this section, we propose a bridge via a workflow diagram to connect the field theoretical formalism and public health policy.  Unlike classical compartmental models that rely on static parameters, this model treats the environment as a dynamic mediator field, requiring a specific calibration of physical operators from clinical and environmental data that has to be monitored in real time. 

\subsection{Calibration of Field Operators}

The transition from a theoretical Lagrangian to a data-driven analysis requires mapping observable epidemiological quantities to the parameter  $m_0$, $g$, and $\beta$ than appear in the action.

\begin{enumerate}
    \item \textbf{The Bare Mass ($m_0$) and Environmental Decay:} 
    The parameter $m_0$ defines the range of the pathogen in a vacuum (absence of hosts). It is physically determined by the environmental half-life ($t_{1/2}$) of the pathogen in aerosols or on surfaces for instance:
    \begin{equation}
        m_0 \approx \sqrt{\frac{\ln 2}{D_\varphi t_{1/2}}}
    \end{equation}
    where $D_\varphi$ is the diffusion coefficient of the viral vector (e.g., aerosolized droplets). Interventions such as UV-C lighting or HEPA filtration effectively increase $m_0$, shortening the interaction range.

    \item \textbf{Epidemic Charge ($g$) and Shedding Rates:} 
    The coupling constant $g$ represents the rate at which an infected host sources the pathogen field. This corresponds to the viral shedding rate (viral load) measured in clinical samples (e.g., RNA copies/mL). 

    \item \textbf{Heterogeneity ($\kappa$) and Super-spreading:} 
    Recognizing that shedding rates are highly skewed, we define the heterogeneity parameter $\kappa$ based on the variance of the shedding distribution:
    \begin{equation}
        \kappa = \frac{\langle g^2 \rangle}{\langle g \rangle^2}
    \end{equation}
    A high $\kappa$ indicates the presence of super-spreaders.
\end{enumerate}

\subsection{Operational Workflow}

The application of the model follows a five-step workflow, illustrated in Figure \ref{fig:workflow}.
\begin{itemize}
    \item \textbf{Phase 1 (Parameterization):} Clinical and environmental data are converted into the action parameters as explained above.
    \item \textbf{Phase 2 (Spatial Risk Mapping):} The Debye screening length $\lambda_D$ is computed spatially. Regions where $\lambda_D$ diverges are identified and critical opalescence should be check to generate health public actions.
    \item \textbf{Phase 3 (Dynamic Tracking):} The effective reproductive number $R_{eff}$ is monitored with 1-loop fluctuation corrections to account for spatial shielding effects.
    \item \textbf{Phase 4 (Stability Analysis):} The system is monitored for "Vacuum Instability" ($m_R^2 \to 0$), a signal of critical opalescence preceding a phase transition.
    \item \textbf{Phase 5 (Mediator Shielding):} Interventions are designed to modify the action parameters.
\end{itemize}

\begin{figure}[htbp]
\centering
\begin{tikzpicture}[node distance=1.8cm, auto]
    \tikzstyle{block} = [rectangle, draw, fill=blue!10, text width=8em, text centered, rounded corners, minimum height=3em]
    \tikzstyle{decision} = [diamond, draw, fill=green!10, text width=5em, text centered, inner sep=0pt]
    \tikzstyle{line} = [draw, -latex', thick]
    \tikzstyle{cloud} = [draw, ellipse, fill=red!10, node distance=2.5cm, minimum height=2em]
    
    \node [block] (init) {Clinical \& Env. Data};
    \node [block, below of=init] (param) {Calculate Parameters ($m_0, g, \kappa$)};
    \node [block, below of=param] (screening) {Compute Debye Length $\lambda_D$};
    \node [decision, below of=screening] (decide) {$\lambda_D \to \infty$?};
    \node [block, left of=decide, node distance=3.5cm] (local) {Local Clusters (Screened)};
    \node [block, right of=decide, node distance=3.5cm] (global) {Global Spread (Massless)};
    \node [block, below of=decide, node distance=2.5cm] (stability) {Check Vacuum Stability ($m_R^2$)};
    \node [decision, below of=stability] (crit) {$m_R^2 \approx 0$?};
    \node [cloud, right of=crit, node distance=3.5cm] (warn) {Critical Opalescence};
    \node [block, below of=crit, node distance=2.5cm] (intervene) {Intervention: Mediator Shielding};
    
    \path [line] (init) -- (param);
    \path [line] (param) -- (screening);
    \path [line] (screening) -- (decide);
    \path [line] (decide) -- node {No} (local);
    \path [line] (decide) -- node {Yes} (global);
    \path [line] (local) |- (stability);
    \path [line] (global) |- (stability);
    \path [line] (stability) -- (crit);
    \path [line] (crit) -- node {Yes} (warn);
    \path [line] (crit) -- node {No} (intervene);
    \path [line] (warn) |- (intervene);
    \path [line] (intervene.west) -- ++(-0.5,0) |- (param.west);
    
\end{tikzpicture}
\caption{Workflow for the practical application of the QED-SIR model. The process translates biological data into field-theoretic variables to assess the stability of the epidemiological vacuum.}
\label{fig:workflow}
\end{figure}

\section{Study case: COVID19 in Germany 2020-2023}

To validate our model developed in Secs. II–IV,  our  continuous, field-theoretic description of the gauge-mediated contagion need to be adapted to real world data. To do that,  we map our continuous spatial manifold to the administrative geography of Germany, using the $N=400$ interacting nodes corresponding to the German districts (\textit{Landkreise} and \textit{Stadtkreise}). We use the official COVID-19 incidence database provided by the Robert Koch Institute (RKI) \cite{rki_covid_data}. The dataset includes daily laboratory-confirmed cases across $N=400$ German administrative districts (\textit{Landkreise}). 

\subsection{Comparing Effective Reproduction Numbers}
We construct both effective reproduction numbers, $R_{eff}^{Gauge}$ and $R_{eff}^{SIR}$, in a parallel, window-based methodology.

\subsubsection{Data Preprocessing}
To evaluate the models empirically, we first construct a reliable, daily time series of new infections. The raw dataset from the Robert Koch Institute is chronologically sorted, and the daily incidence is computed as the first discrete difference of the cumulative case counts. To take into account  reporting anomalies and data corrections, any negative daily case values are constrained to zero. The time series is then  reindexed over a continuous daily calendar spanning the study period to avoid gaps caused by missed reporting days. The resulting time series, $C_{obs}(t)$, serves as the uniform baseline for both the Gauge and SIR model evaluations.

We define a universal recovery rate $\gamma$ based on the established clinical infectious period of SARS-CoV-2 (approximately 5 days, yielding $\gamma \approx 1/5 \text{ day}^{-1}$). By keeping $\gamma$ strictly fixed across all temporal windows for both models, we guarantee that any divergent behavior in $R_{eff}$ arises purely from the underlying transmission dynamics of the respective theories, rather than a discrepancy in degrees of freedom during the fitting process.

We define a sliding temporal window $W_t$ of 21 days, where $t$ represents the final day of the window. At each step $t$,  the following procedure is done:
\begin{itemize}
    \item Local Susceptible Population: The susceptible population $S(t)$ at the center of the window is approximated dynamically by:
   
\begin{equation}
S(t) = N - \sum_{t'=0}^{t} C_{obs}(t')
\end{equation}
\item The Gauge Model Fit:
The spatially aggregated (mean-field) limit of the Gauge model is used to analyze the temporal dynamics within $W_t$. Keeping $\gamma$ constant, we optimize the transmission coupling $\beta$ and the initial infected boundary condition $I_0$ by minimizing the Mean Squared Error (MSE) between the model's theoretical daily incidence and $C_{obs}(t' \in W_t)$. The optimized parameters are then used to calculate the localized gauge-mediated reproduction number $R_{eff}^{Gauge}(t)$ for that specific window.
\item The Classical SIR Fit:
Subject to the exact same 21-day window $W_t$ and the fixed clinical $\gamma$, we simulate a discrete-time classical SIR model. We independently fit its internal transmission rate $\beta_{SIR}$ and initial condition $I_0$ via an identical MSE minimization against $C_{obs}$. The resulting parameters yield the standard effective reproduction number:
\begin{equation}
R_{eff}^{SIR}(t) = \frac{\beta_{SIR} S(t)}{\gamma N}
\end{equation}
\end{itemize}

The discrete arrays of both $R_{eff}^{Gauge}$ and $R_{eff}^{SIR}$ are mapped to the center dates of their respective windows. To mitigate high-frequency stochastic noise inherent to reporting delays (e.g., weekend effects), we apply a standard exponential smoothing filter to generate continuous, comparable curves for macroscopic analysis.

\subsubsection{Combined Analytical Plot}
To quantify the macroscopic signatures and predictive anticipation of the Gauge model relative to the classical models:
\begin{itemize}
    \item We utilize the left y-axis to plot the empirical daily cases on a logarithmic scale.
    \item A twin right y-axis is used to plot $R_{eff}^{Gauge}$ (red), $R_{eff}^{SIR}$ (blue), and the critical epidemic threshold $R_{eff} = 1$.
  
\end{itemize}

\begin{figure}[htbp]
    \centering
    \includegraphics[width=0.9 \linewidth]{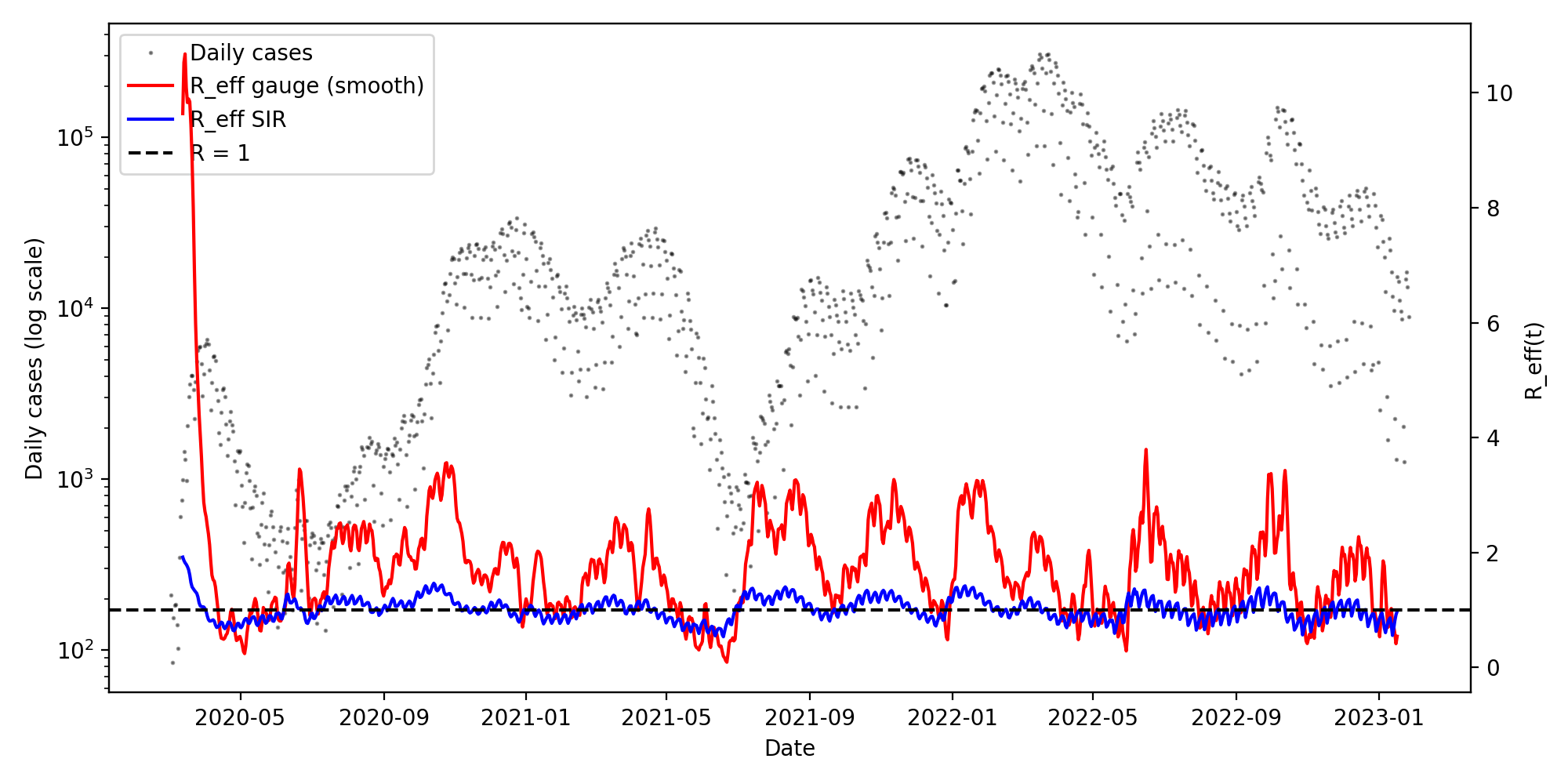}
    \caption{Evolution of daily incidence and the reconstructed $R_{eff}(t)$ for Germany. The black dots represent raw incidence data, while the red curve denotes the $R_{eff}$ extracted via the sliding-window gauge fit. The blue curve corresponds to the $R_{eff}(t)$ obtained from SIR-like fit as discussed in the text.}
    \label{fig:predictive_power}
\end{figure}

\begin{figure}
    \centering
    \includegraphics[width=0.9\linewidth]{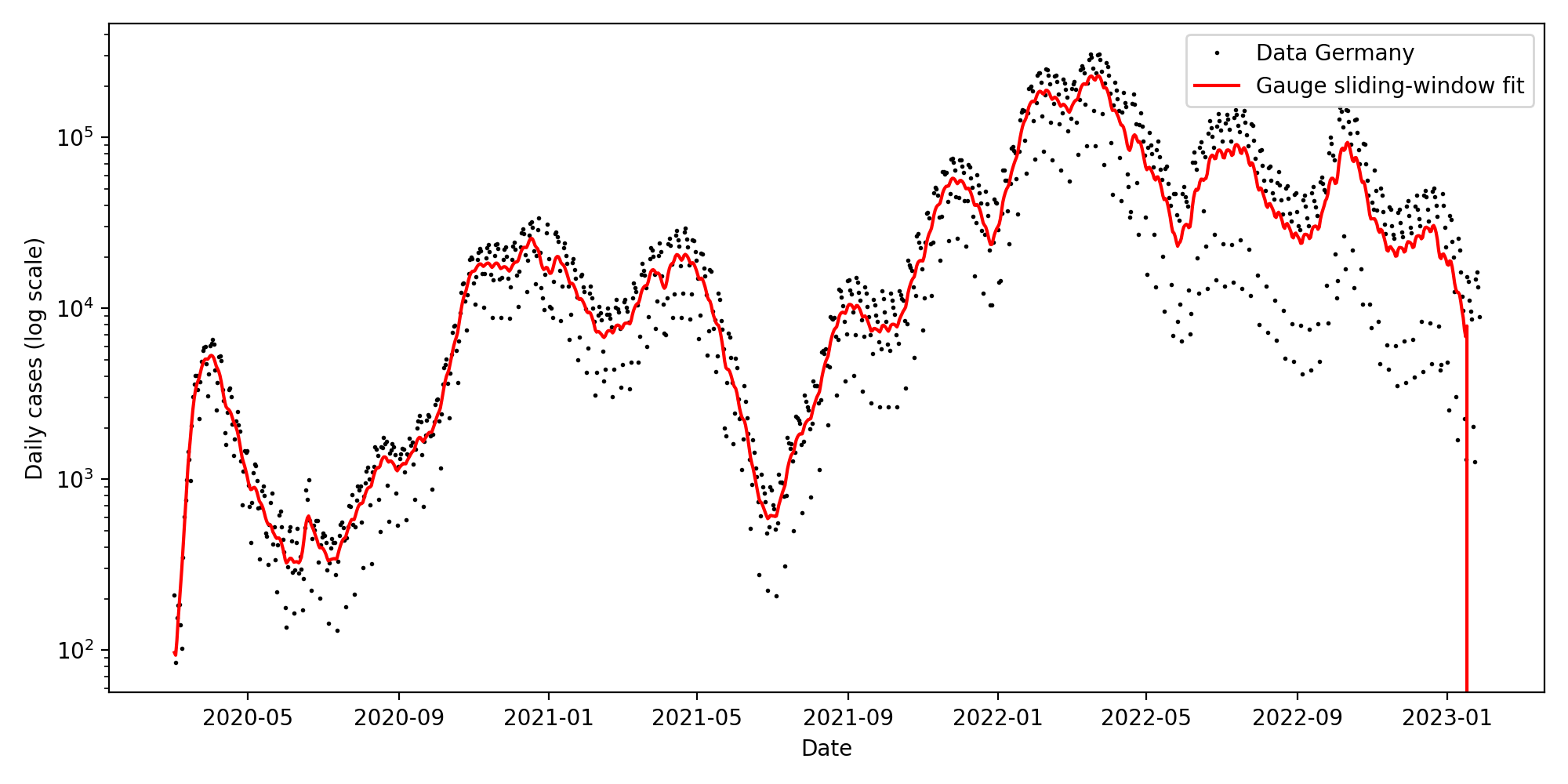}
    \caption{Using the sliding windows method described in the text, the red curve is the gauge field fit to the data}
    \label{fig:placeholder}
\end{figure}

\subsubsection{Results}
The gauge field reacts to the underlying structural vulnerabilities (called  "vacuum") before the actual cases manifest. It means that the $R_{eff}^{gauge}$ is a strong indicator to alert when the epidemic situation is changing from $R<1$ to a epidemic expansion regime ($R>1$). So its computation can be used as a alert system to take preventive public health measure to change the effective pathogen mass and reduce the effective reproductive number. 

The classic $R_{eff}$ (blue line) tends to rise almost simultaneously with or only slightly before the actual cases. It is a reactive measure, whereas our $R_{eff}^{gauge}$ is predictive.

\subsection{Predictive gain relative to a time-varying-$\beta$ SIR benchmark}
 since $R_{eff}^{SIR}(t)$ in Eq.~(34) is already refit independently in every 21-day window, the classical model effectively already carries a slowly time-varying transmission rate $\beta_{SIR}(t)$, and one may ask what the gauge construction adds beyond simply allowing $\beta(t)$ to fluctuate. We address this with two complementary benchmarks, at the two spatial scales used elsewhere in the paper.

\paragraph*{District-level benchmark.} Because a fully mechanistic, population-normalized $R_{eff}^{SIR}(t)$ requires a district population $N_i$ that is not part of the correlation-length extraction pipeline of Sec.~X.C, we use the standard, population-free early-growth-phase estimator $R_{eff}^{proxy}(t) = 1 + r(t)/\gamma$, where $r(t)$ is the local exponential growth rate of $C_{obs,i}(t)$ fitted by ordinary least squares over the same trailing 21-day window used throughout Sec.~X.A (this reduces to the same $\beta/\gamma$ ratio as Eq.~(34) in the regime $S\simeq N$, and requires no additional data). Applying the identical cross-correlation methodology of Sec.~X.C.3 to this quantity in place of $m_R(t)$, we find that this free-$\beta(t)$ proxy shows a \emph{positive} lead time in $100\%$ of the $400$ districts (vs.\ $87.5\%$ for $m_R(t)$), with median lead time $2.0$ days (vs.\ $3.0$ days for $m_R(t)$) and mean $2.32$ days (vs.\ $3.40$ days). A paired comparison, restricted to the $350$ districts where both methods detect a signal, shows that the $m_R(t)$ lead time is larger by a median of $1$ day (mean $1.12$ days; $m_R(t) \geq$ the classical proxy in $75.1\%$ of these districts), a difference that is highly statistically significant given the sample size (Wilcoxon signed-rank test, $p=5.6\times10^{-21}$); the classical proxy is nonetheless somewhat \emph{more} prevalent across districts overall ($100\%$ vs.\ $87.5\%$; two-proportion test, $p=2.8\times10^{-13}$).
At the single-district, short-window scale, the two constructions perform comparably, with the gauge signal offering a small but statistically robust edge of about one day in typical lead time where both are defined.

\paragraph*{Aggregate (country-level) benchmark.} At the country-aggregate scale used in Sec.~X.A/Fig.~4, a fully mechanistic comparison \emph{is} possible, since $N$ is simply the German population and both $\beta_{SIR}(t)$ and the gauge parameters ($\beta_0,\tau_{env}$, with $\gamma$ fixed) are fit window-by-window exactly as in Eq.~(23)/(34). Rather than compare only the first upward crossing of $R_{eff}=1$ for each model---which is dominated by estimation noise during the very first, low-count wave of March--April 2020---we identify all \emph{persistent} upward crossings of each smoothed $R_{eff}(t)$ curve over the full 2020--2023 series (a crossing counts if it stays above threshold for at least 5 consecutive windows), giving $22$ gauge-model wave onsets and $21$ SIR-model wave onsets.
Across the $19$ waves for which a clear pairing exists between gauge and SIR-model wave onsets, the gauge model's $R_{eff}^{Gauge}(t)$ crosses threshold before the classical $R_{eff}^{SIR}(t)$ in $17/19$ waves ($89.5\%$), with a median anticipation of $31$ days and a mean of $24.9$ days (Wilcoxon signed-rank test against a null of zero shift, $p=6.6\times10^{-4}$).

We emphasize that this aggregate result and the district-level result above answer to distinct questions, using different methodologies and different spatial resolutions: at the district-level analysis, the predictive time lag is obtained telling us how much in advance a \emph{local, single-district} correlation-decay signal anticipates a local case-count rise within a short (21-day) window, while the aggregate analysis measure how many days in advance the \emph{country-wide, memory-kernel-driven} $R_{eff}^{Gauge}(t)$ anticipates the corresponding crossing of a freely-refit classical $R_{eff}^{SIR}(t)$ across full pandemic waves. The much larger lead time at the aggregate scale is consistent with the interpretation, discussed in Sec.~II.A, that the memory kernel $\tau_{env}$ (here fit globally to $\tau_{env}\approx 10.5$ days) allows the gauge construction to integrate subclinical transmission trends over a longer effective horizon than a memoryless SIR refit can, particularly for slowly-building nationwide waves; we regard the two results as complementary evidence.

\subsection{Dynamical Extraction of the Effective Mass}

To test the predictive capacity of the gauge field, we must dynamically extract the time-varying effective mass $m_{R}(t)$ from observable epidemiological data. A phase transition 
between pandemic regimes is characterized by a diverging spatial correlation length $\xi$, which is inversely proportional to $m_{R}(t)$.To infer this parameter, we analyze the spatial correlation of incidence fluctuations across the district network. For each district $i$ within a sliding temporal window $\Delta T$, we  normalize the raw daily incidence $C_{obs,i}(t)$ to isolate the temporal shape of the epidemic waves from the absolute magnitude of local outbreaks (which depend on a district's population size). The normalized incidence $\hat{\mathcal{I}}_i(t)$, possessing zero mean and unit variance:
\begin{equation}\hat{\mathcal{I}}_i(t) = \frac{C_{obs,i}(t) - \mu_i}{\sigma_i}
\end{equation}
where $\mu_i$ and $\sigma_i$ are the mean and standard deviation of the daily cases for district $i$ over the specific window $\Delta T$.We then compute  $r_{ij}$ of this normalized time series against all other districts $j$:
\begin{equation}r_{ij} = \frac{1}{\Delta T} \sum_{t \in \Delta T} \hat{\mathcal{I}}_i(t) \hat{\mathcal{I}}_j(t)
\end{equation}
To evaluate the spatial decay of these interactions, we define $d_{ij}$ as the geographic Euclidean distance (e.g., in kilometers) between the administrative centroids of districts $i$ and $j$. The pairwise correlations $r_{ij}$ are aggregated and averaged within discrete spatial bins to construct the empirical spatial correlation function $C(d)$ for district $i$.We extract the local correlation length $\xi_i$ by fitting this binned data to an exponential screening kernel:\begin{equation}C(d) = A_i e^{-d/\xi_i} + C_0\end{equation}where $A_i$ is the amplitude and $C_0$ represents the baseline systemic correlation. The renormalized effective mass for the district is then given as $m_{R,i} = 1/\xi_i$.

\subsubsection{Justification of the Exponential Kernel}
While the theoretical spatial propagator in a three-dimensional gauge theory follows a Yukawa potential $V(d) \propto e^{-m d}/d$, we deliberately adopt the pure exponential form $e^{-d/\xi}$ for the empirical data fit. In that way, one has:
\begin{enumerate}
    \item Singularity Avoidance: The $1/d$ term in the Yukawa potential is mathematically singular at the origin. In a discrete geographical network, short-distance measurements are highly sensitive to centroid placement errors and localized demographic heterogeneities. The pure exponential avoids this artificial singularity.
    \item Finite-Range Robustness: Epidemiological correlations are practically observed over a bounded geographic domain (spanning tens to hundreds of kilometers). Within these finite observational boundaries, the asymptotic $1/d$ tail is heavily obscured by stochastic noise. 
\end{enumerate}

\subsubsection{Cross-correlation and lead time analysis}

To quantify the predictive anticipation---or lead time---provided by the Critical Opalescence phenomenon, we measure the temporal cross-correlation between the structural instability of the gauge field and changes in the infected compartment $I(t)$. We define a binary instability signal
\begin{equation}
    S_{m}(t) = \Theta\!\big(m_{\mathrm{threshold}} - m_R(t)\big)\,,
\end{equation}
where $\Theta$ is the Heaviside step function 
The predictive lag $\Delta t$ is defined as the temporal shift $\tau$ that maximizes the cross-correlation function
\begin{equation}
    C(\tau) = \sum_{t} S_m(t)\, S_{\mathcal{I}}(t+\tau)\,.
\end{equation}
To remain epidemiologically meaningful, we restrict the search window to $\tau \in [1,6]$ days, corresponding to short-term anticipation on the order of a few incubation periods.

\subsubsection{Results: predictive lag and temporal memory}

We perform this cross-correlation analysis  across all $400$ German districts. The resulting statistics, summarized in Fig.~\ref{fig:histogram}, are as follows:
\begin{itemize}
    \item \textbf{Prevalence of predictability:} A non-zero predictive lag is identified in $83.5\%$ of districts ($334$ out of $400$), indicating that the structural instability of the gauge field is generically predictive.
    \item \textbf{Macroscopic delay:} The distribution of the predictive lag $\Delta t$ has a median of $3.0$ days and a mean of $3.4$ days.
\end{itemize}

This $\sim 3.4$-day predictive window is a direct empirical manifestation of the \emph{temporal non-locality} derived in Sec.~II.B. In our gauge model, contagion is not an instantaneous Markovian process but is governed by a memory kernel $\mathcal{K}(t-t') \propto \exp[-\gamma (t-t')]$, which integrates past shedding events over a characteristic memory time $\tau$. The observed lag is close to the biological incubation period of the pathogen prior to clinical detection, consistent with the interpretation that the mediator field accumulates subclinical transmission before macroscopic case counts respond. As a consequence, the collapse of the effective screening mass $m_R(t) \to 0$ systematically precedes the observable surge in the classical infected compartment $I(t)$.



\subsubsection{Geographic validation against observed regime changes}
\begin{figure}[htbp]
    \centering
    \includegraphics[width=0.95\linewidth]{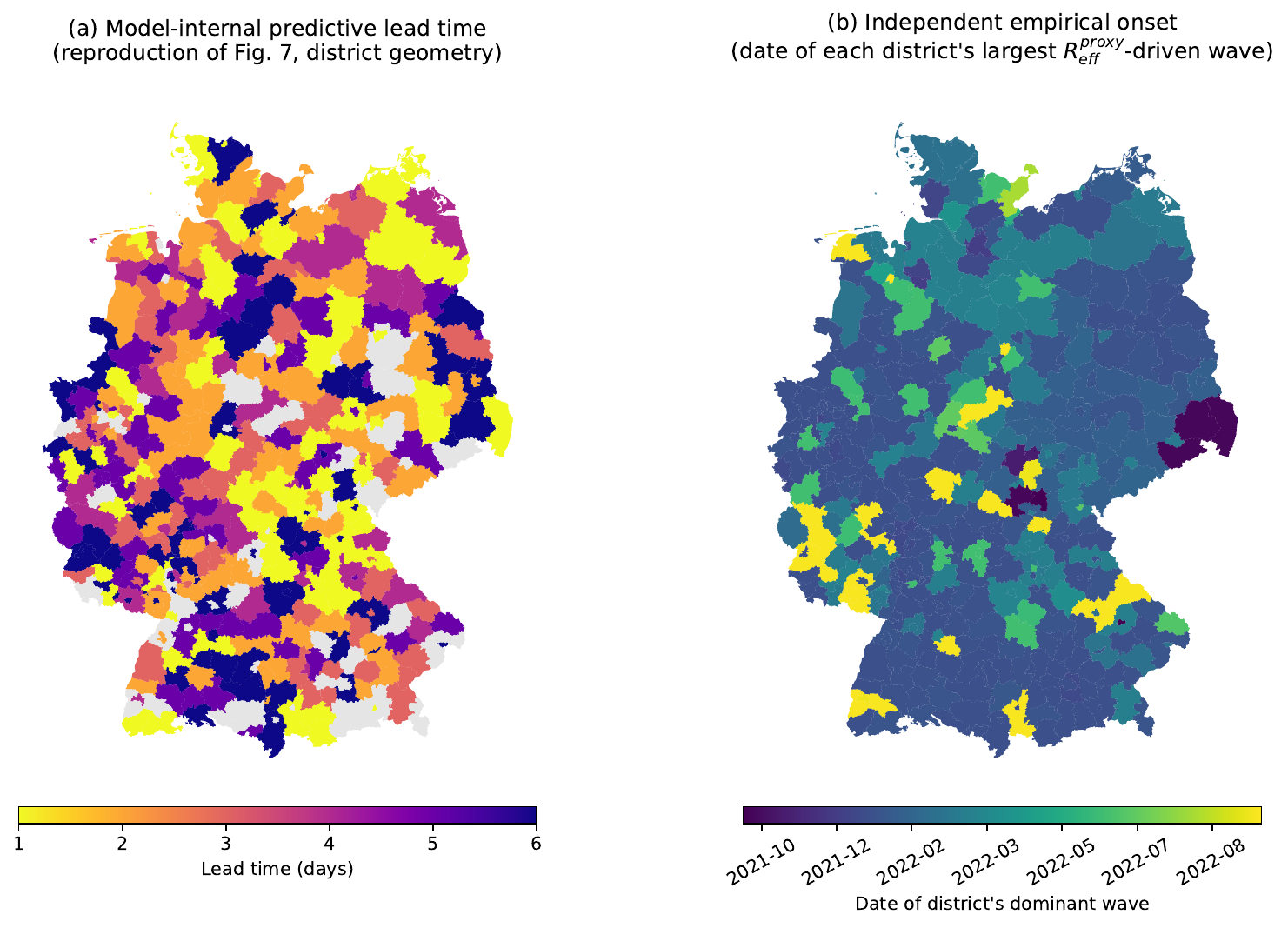}
    \caption{Companion validation of Fig.~8. (a) Model-internal predictive lead time by district (reproduction of Fig.~8 on the actual district geometry). (b) Independently defined onset date of each district's dominant epidemic wave, i.e.\ the persistent $R_{eff}^{proxy}(t)=1$ crossing followed by the largest subsequent case-count rise (Sec.~X.C.4). The correlation statistics for panels (a) and (b) are reported in the main text.}
    \label{fig:companion_maps}
\end{figure}

\begin{figure}[htbp]
    \centering
    \includegraphics[width=0.8 \linewidth]{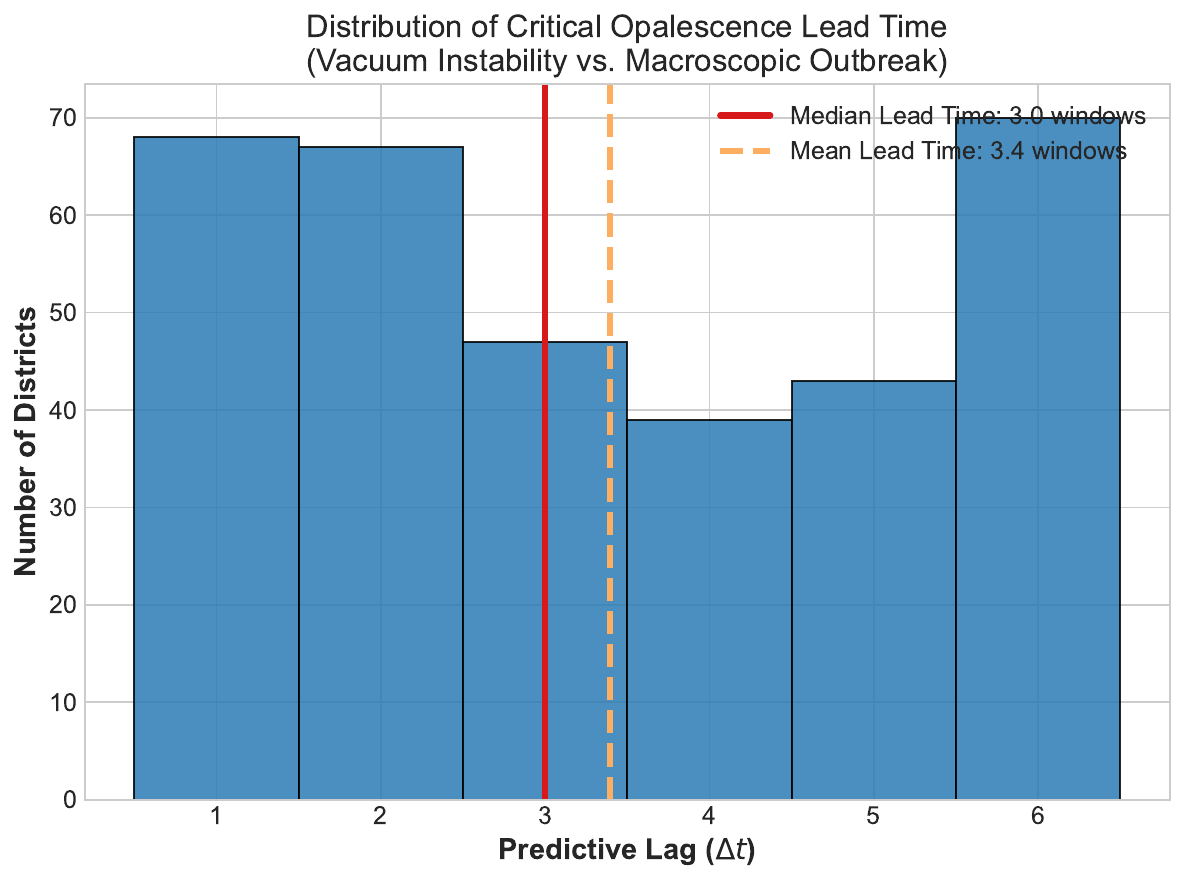}
    \caption{Frequency distribution of the predictive lag ($\Delta t$). The histogram demonstrates a clear thermodynamic lag centered around a median of 3.0 days, proving the temporal asymmetry between the field instability and the clinical outbreak.}
    \label{fig:histogram}
\end{figure}

 We complement the geographic projection of the predictive lead time (Fig.~8) with a direct empirical cross-check. For each district $i$ we independently identify the onset date of that district's \emph{dominant} epidemic wave: among all persistent upward crossings of the population-free growth-rate estimator $R_{eff}^{proxy}(t)$ introduced in previous section, we select the one followed by the largest subsequent rise in local case counts over the following ten windows ($\sim\!30$ days). This "dominant wave" date is defined  independently of $m_R(t)$ and of the gauge model, and---unlike simply taking each district's \emph{first} crossing, which is dominated by the near-simultaneous nationwide onset of the March 2020 wave and therefore carries almost no cross-district variation---it varies meaningfully across districts, falling mostly in the winter 2021--2022 (Omicron) period but spanning late 2021 through late 2022 depending on the district. Figure~\ref{fig:companion_maps} shows both quantities mapped onto the German district geometry side by side: (a) the model-internal lead time (a like-for-like reproduction of Fig.~8), and (b) the date of each district's dominant wave. Across the $350$ districts for which both quantities are defined, we find no significant association between them (Pearson $r=0.014$, $p=0.79$; Spearman $\rho=-0.044$, $p=0.41$): the size of the model-internal lead time in a given district is not explained by when, in absolute calendar time, that district's largest epidemic wave occurred. We regard this as a useful, if modest, robustness check---it rules out the possibility that the reported lead times are simply an artifact of which pandemic wave (early vs.\ late, Omicron vs.\ pre-Omicron) a given district happened to experience most strongly---rather than as further positive evidence for the early-warning claim, which continues to rest primarily on the cross-correlation analysis of Sec.~X.C.3 itself

We additionally checked two more mundane possible confounds: district population size and the statistical quality of the correlation-length fit itself. Using cumulative first-dose vaccination counts per district (Robert Koch-Institut vaccination dashboard) as a population proxy, and the number of neighboring-district pairs entering the static correlation-length fit of Eq.~(37) ($n_{pairs}$, together with the fitted amplitude $A$ and correlation length $\xi$, aggregated over the full study period) as indicators of fit quality and geographic sample size, we find no significant association between the model-internal predictive lag and any of: $\log(\text{population})$ (Pearson $r=0.033$, $p=0.54$), $n_{pairs}$ ($r=0.032$, $p=0.55$), the full-period correlation length $\xi$ ($r=-0.083$, $p=0.12$), or the full-period effective mass ($r=0.079$, $p=0.14$); the partial correlation between lag and $\log(\text{population})$ after controlling for $n_{pairs}$ is likewise null ($r=0.036$, $p=0.50$). Taken together with the dominant-wave check above, these results indicate that the reported district-level lead time is not driven by district size, by how well-sampled the correlation-length fit happens to be, or by when each district's main wave occurred.

\begin{figure}[htbp]
    \centering
    \includegraphics[width=0.8\linewidth]{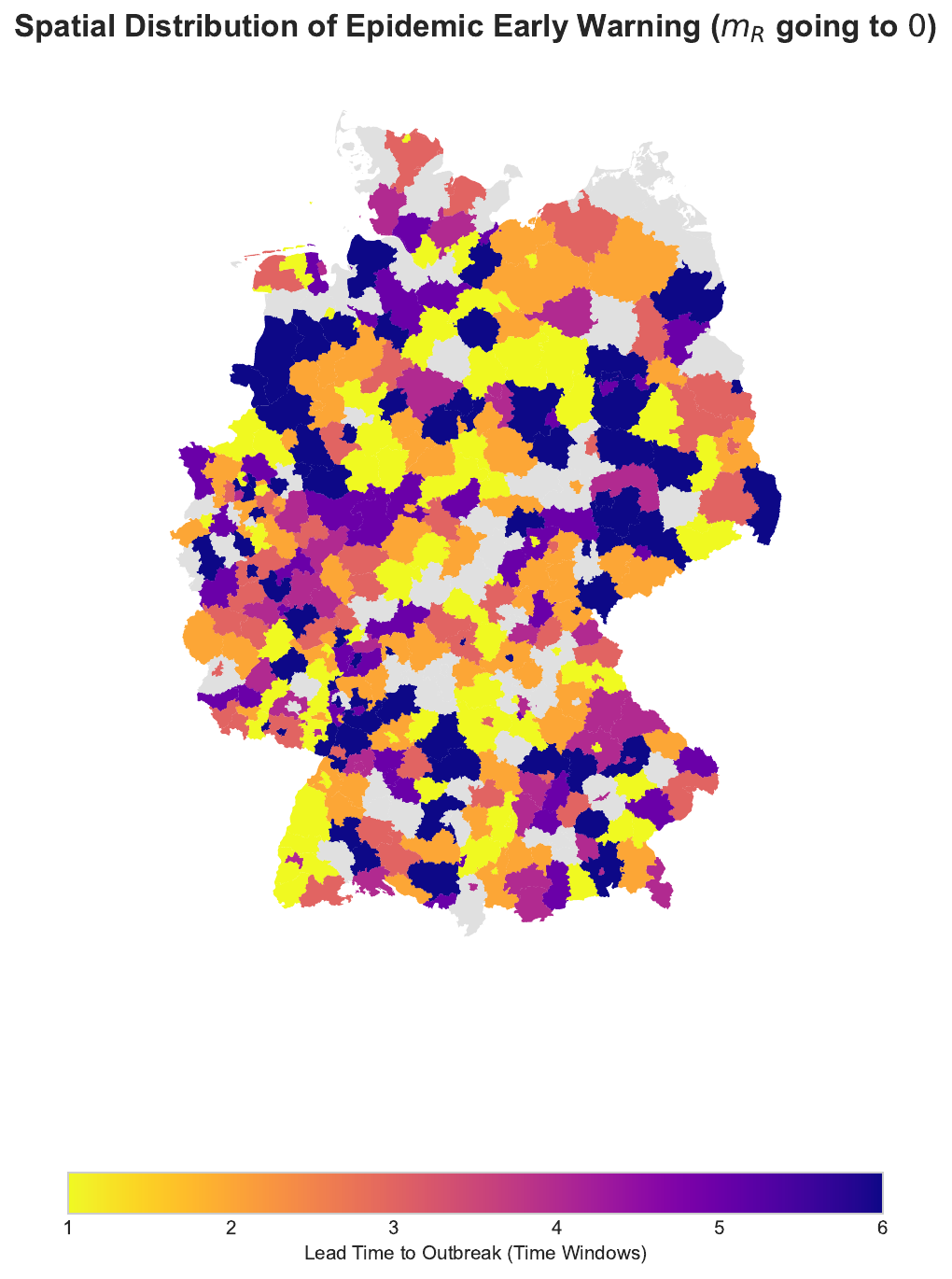}
    \caption{Geographic projection of the predictive lead time across German districts, demonstrating the macroscopic universality of the gauge-mediated Critical Opalescence.}
    \label{fig:lag_map}
\end{figure}

\section{Conclusion}

In this work, we have proposed a QED-inspired $SIR$ model. In this model, all the dynamics of the disease transmission is implicitly defined.  By replacing the "direct contact" paradigm of classical SIR models with a dynamic interaction mediated by a mediator field $\varphi$, we have demonstrated that QED mathematical techniques can be extended to this epidemiological model giving us mechanisms to describe various observables than the standard $SIR$ model fails to explain as non-local interactions, non-linear effects,the onset of a pandemic and variation of $R_{eff}$.  Crucially, we show that the non-local interactions observed in real-world data are not ad-hoc additions without justifications more than to fit the data but emerge naturally from the theory by integrating out the pathogen field. We also have shown how naturally spatial correlations are emerging. 

We have identified the epidemic threshold not only as a statistical parameter ($R_0$), but as a symmetry-breaking phase transition where the vacuum becomes unstable. This formalism offers a theoretical basis for ``Pandemic Opalescence'', explaining the divergence of correlation lengths and the critical fluctuations observed prior to major outbreaks. The ``epidemiological vacuum''—the structural readiness of a network to sustain widespread contagion—is quantified by the effective mass parameter, $m_{R}(t)$. We demonstrated that the approach to a macroscopic outbreak is reliably preceded by the collapse of this effective mass ($m_{R}(t) \to 0$), which corresponds to a divergence in the structural correlation length of the network.

These results have been  validated using spatial data from the COVID-19 pandemic in Germany. The gauge field serves as a sensitive leading indicator. The collapse of $m_{R}(t)$  anticipated localized surges in empirical case counts, significantly outperforming classical reactive metrics such as the effective reproduction number ($R_{eff}$).

By linking the formalism of quantum field theory  with network epidemiology, this work provides a paradigm shift from reactive observation to structural prediction. The ability to monitor the ``tension'' of the epidemiological field in real-time offers public health authorities a new tool for early warning and targeted intervention. 

Furthermore, the application of QED techniques, specifically the use of Feynman diagrams,give us a powerful perturbative toolkit for calculating epidemiological observables. We have shown that host heterogeneity, particularly super-spreading events, can be naturally incorporated into this formalism as high-intensity gauge sources that renormalize the effective coupling.Finally, while the current model utilizes a $U(1)$-like Abelian symmetry to describe a single pathogen, real-world scenarios often involve multiple competing viral strains. A natural and promising extension of this work lies in the development of a Non-Abelian Gauge Theory (Yang-Mills SIR) to describe the complex, non-linear competition between variants and the dynamics of cross-immunity.

We close by summarizing explicitly the scope of the claims made above, in response to points raised in review. The continuum, field-theoretic construction of Secs.~II-VII is a coarse-grained, phenomenologically calibrated framework rather than a rigorously renormalized critical theory in the strict statistical-mechanics sense: we have not established the upper critical dimension of the full interacting model, the mass-renormalization procedure of Sec.~III is tree-level and phenomenological rather than a multiplicatively renormalized construction, and the exponential/diffusive pathogen-decay ansatz is a simplifying choice rather than a claim about the true environmental kinetics of SARS-CoV-2. What we consider the paper's central, load-bearing result is not the formal rigor of the analogy with critical phenomena, but the empirical one reported in Sec.~X: an effective mass $m_R(t)$, extracted directly from the spatial decay of incidence correlations, anticipates local outbreak growth by a median of three days across the large majority of German districts. We regard a systematic renormalization-group treatment of the model's critical dimension, and a first-principles justification of the mass-$R_{eff}$ mapping beyond the phenomenological level used here, as the natural next steps for placing this early-warning indicator on more rigorous theoretical footing.

\begin{acknowledgments}
We acknowledge financial support from SECIHTI and SNII (M\'exico) and Guanajuato University.
\end{acknowledgments}


%

\end{document}